\documentclass[pra,twocolumn,showpacs]{revtex4-1}
\usepackage{amsmath,amscd,amsfonts,amssymb,color}
\usepackage{graphicx,amsfonts,epsf}
\begin{document}
\title{Experimental protection against evolution 
of states in a subspace via a super-Zeno 
scheme on an NMR quantum information processor}
\author{Harpreet Singh}
\email{harpreetsingh@iisermohali.ac.in}
\affiliation{Department of Physical Sciences, Indian
Institute of Science Education \& 
Research (IISER) Mohali, Sector 81 SAS Nagar, 
Manauli PO 140306 Punjab India.}
\author{Arvind}
\email{arvind@iisermohali.ac.in}
\affiliation{Department of Physical Sciences, Indian
Institute of Science Education \& 
Research (IISER) Mohali, Sector 81 SAS Nagar, 
Manauli PO 140306 Punjab India.}
\author{Kavita Dorai}
\email{kavita@iisermohali.ac.in}
\affiliation{Department of Physical Sciences, Indian
Institute of Science Education \& 
Research (IISER) Mohali, Sector 81 SAS Nagar, 
Manauli PO 140306 Punjab India.}
\begin{abstract}
We experimentally demonstrate the freezing of evolution of
quantum states in one- and two-dimensional subspaces of two
qubits, on an NMR quantum information processor. State
evolution was frozen and leakage of the state from its
subspace to an orthogonal subspace was successfully
prevented using super-Zeno sequences~\cite{dhar-prl-06}, 
comprising of a set of
radio frequency (rf) pulses punctuated by
pre-selected time intervals.  We demonstrate the efficacy of
the scheme by preserving different types of states,
including  separable and maximally entangled states in one-
and two-dimensional subspaces of two qubits.  The change in
the experimental density matrices was tracked by carrying
out full state tomography at several time points.  
We use the fidelity measure for the
one-dimensional case and the leakage (fraction) into the
orthogonal subspace for the two-dimensional case,
as qualitative indicators to estimate the resemblance of the
density matrix at a later time to the initially prepared
density matrix.
For the
case of entangled states, we additionally compute an
entanglement parameter to indicate the presence of
entanglement in the state at different times.  
We experimentally demonstrate that 
the super-Zeno scheme is able to
successfully confine state evolution to the one- or two-dimensional
subspace being protected.
\end{abstract}
\maketitle
\section{Introduction}
\label{intro}
Using frequent measurements to project a quantum system back
to its initial state and hence slow down its time evolution
is a phenomenon known as the quantum Zeno
effect~\cite{sudarshan,facchi-jmp-10,facchi-jphyconf-09}. If
the measurements project the system back into a
finite-dimensional subspace that includes the initial state,
the state evolution remains confined within this subspace
and the subspace can be protected against leakage of
population using a quantum Zeno
strategy~\cite{facchi-prl-02,busch-jphyconf-10}.  An
operator version of this phenomenon has also been suggested
recently~\cite{wang-prl-13,li-pra-13}.  Zeno-like schemes
have been used for error prevention~\cite{erez-pra-04}, and
to enhance the entanglement of a state and bring it to a
Bell state, even after entanglement sudden
death~\cite{sabrina-prl-08,oliveira-pra-08}.  It has been
shown that under certain assumptions, the Zeno effect can be
realized with weak measurements and can protect an unknown
encoded state against environment
effects~\cite{silva-prl-12}.  An interesting  quantum
Zeno-type strategy for state preservation, achieved using a
sequence of non-periodic short duration pulses,
has been
termed the super-Zeno scheme~\cite{dhar-prl-06}. The
super-Zeno scheme does not assume any Hamiltonian symmetry,
does not involve projective quantum measurements,
and achieves a significant betterment of the leakage
probability as compared to standard Zeno-based preservation
schemes. 
Similar schemes involving dynamical decoupling
have been devised to suppress qubit pure dephasing and
relaxation~\cite{uhrig-njp-08,yang-prl-08}.  Another
scheme to preserve entanglement in a two-qubit spin-coupled
system has been constructed, which unlike the
super-Zeno scheme, is based on a sequence of
operations performed periodically on the system  in a given
time interval~\cite{hou-annal-12}.

There are several experimental implementations of the
quantum Zeno phenomenon, including suppressing
unitary evolution driven by external fields between
the two states of a trapped ion~\cite{itano-pra-90}, in
atomic systems~\cite{bernu-prl-08}
and suppressing failure events in a linear optics quantum
computing scheme~\cite{franson-pra-04}. Decoherence control
in a superconducting qubit system has been proposed using
the quantum Zeno effect~\cite{tong-pra-14}.  
Unlike the super-Zeno and dynamical decoupling schemes that
are based on unitary pulses, the quantum Zeno effect
achieves suppression of state evolution using
projective measurements.
The quantum
Zeno effect was first demonstrated in NMR by a set of
symmetric $\pi$ pulses~\cite{xiao-pla-06}, wherein  
pulsed magnetic field gradients and controlled-NOT
gates were used to mimic
projective measurements.
The
entanglement preservation of a Bell state in a two-spin
system in the presence of anisotropy was demonstrated using
a preservation procedure involving free evolution and
unitary operations~\cite{manu-pra-14}.  An NMR scheme to
preserve a separable state was constructed using the
super-Zeno scheme and the state preservation was found to be
more efficient as compared to the standard Zeno
scheme~\cite{ting-chinese-09}.  
The quantum Zeno effect was used to stabilize superpositions
of states of NMR qubits against dephasing, using an
ancilla to perform the measurement~\cite{kondo-qph-14}.
Entanglement preservation
based on a dynamic quantum Zeno effect was demonstrated
using NMR wherein frequent measurements were implemented
through entangling the target and measuring
qubits~\cite{zheng-pra-13}.

This work focuses on two applications of the super-Zeno
scheme: (i) Preservation of a state by freezing state
evolution (one-dimensional subspace protection) and (ii)
Subspace preservation by preventing leakage of population to
an orthogonal subspace (two-dimensional subspace
protection).  Both kinds of protection schemes are
experimentally demonstrated on separable as well as on
maximally entangled two-qubit states.  
One-dimensional subspace protection is demonstrated
on the separable $\vert 1 1 \rangle$ state and on
the maximally entangled 
$\frac{1}{\sqrt{2}}(\vert 0 1 \rangle - \vert 1 0 \rangle)$
(singlet) state.
Two-dimensional subspace preservation
is demonstrated by choosing the $\{\vert 0 1 \rangle, \vert
1 0 \rangle\}$ subspace in the four-dimensional Hilbert
space of two qubits, and implementing the super-Zeno
subspace preservation protocol on three different states,
namely $\vert 0 1 \rangle$, $\vert 1 0 \rangle$ and
$\frac{1}{\sqrt{2}}(\vert 0 1 \rangle - \vert 1 0 \rangle)$
(singlet) states.  Complete state tomography is utilized to compute
experimental density matrices at several time points.  State
fidelities at these time points were computed to evaluate
how closely the states resemble the initially prepared
states, with and without super-Zeno protection.  The success
of the super-Zeno scheme in protecting states in the
two-dimensional subspace spanned by $\{\vert 0 1 \rangle,
\vert 1 0 \rangle \}$ is evaluated by computing a leakage
parameter, which computes leakage to the orthogonal subspace
spanned by $\{\vert 0 0 \rangle, \vert 1 1 \rangle \}$.  For
entangled states, an additional entanglement parameter is
constructed to quantify the residual entanglement in the
state over time.
State fidelities, the leakage parameter and the entanglement
parameter are plotted as a function of time, to quantify the
performance of the super-Zeno scheme.

The material in this paper is organized as
follows:~Sec.~\ref{theory} gives a concise description of
the theoretical super-Zeno scheme, in Sec.~\ref{expt}
and the subsections therein
we describe the main experimental results, namely freezing
the evolution of a separable and an entangled state and the
prevention of leakage of population from a subspace, both
schemes being implemented on a two-qubit NMR information
processor. Sec.~\ref{concl} contains some concluding
remarks.

\section{The super-Zeno scheme}
\label{theory}
The super-Zeno algorithm to preserve quantum states has
been developed along lines similar to bang-bang control
schemes, and limits the quantum system's evolution to
a desired subspace using a series of 
unitary kicks~\cite{dhar-prl-06}.

A finite-dimensional Hilbert space $\cal{H}$ can be written
as a direct sum of two orthogonal subspaces $\cal{P}$ and
$\cal{Q}$. The 
super-Zeno scheme involves a unitary kick {\bf J},
which can be constructed as 
\begin{equation}
{\bf J = Q-P} 
\label{Jeqn}
\end{equation}
where $\bf P, Q$ are the projection operators
onto the subspaces $\cal{P},\cal{Q}$ respectively.
The action of this specially crafted pulse 
${\bf J}$ on a state
$\vert \psi \rangle \in \cal{H}$
is as follows: 
\begin{eqnarray}
{\bf J} \vert \psi \rangle &=& - \vert \psi \rangle, 
\quad \vert \psi \rangle \in {\cal P} \nonumber \\
{\bf J} \vert \psi \rangle &=&  \vert \psi \rangle, 
\quad \vert \psi \rangle \in {\cal Q} 
\end{eqnarray}
where  ${\cal P}$ is the subspace being preserved.

The total super-Zeno sequence for $N$ pulses is given by
\begin{equation}
W_N(t) = U(x_{N+1}t) {\bf J} \dots {\bf J} U(x_2 t) 
{\bf J} U(x_1 t)
\label{superzenoeqn}
\end{equation}
where $U$ denotes unitary evolution under
the system Hamiltonian and $x_i t$ is the time interval between the
$i$th and $(i+1)$th pulse.
The sequence $\{ x_i t\}$ of time intervals between pulses
is optimized such that if the system
starts out in the subspace $\cal P$, 
after measurement the probability of finding the
system in the orthogonal subspace $\cal Q$ is minimized. 
In this work we
used four inverting pulses interspersed with 
five inequal time intervals in each repetition
of the preserving super-Zeno sequence. The
optimized sequence 
is given by
$\{ x_i \} = \{ \beta, 1/4, 1/2-2\beta, 1/4,
\beta \}$ with $\beta = (3 - \sqrt{5})/8,
i=1 \dots 5$ and $t$ is a fixed time interval
(we use the $x_i$ as worked out in Ref.~\cite{dhar-prl-06}).

The explicit form of the unitary kick ${\bf J}$ depends
on the subspace that needs to be preserved, and in the
following section, we implement 
several illustrative examples for both separable
and entangled states embedded in one- and two-dimensional
subspaces of two qubits.
\section{Experimental demonstration of 
super-Zeno strategies}
\label{expt}
\begin{figure}[t]
\includegraphics[angle=0,scale=1.0]{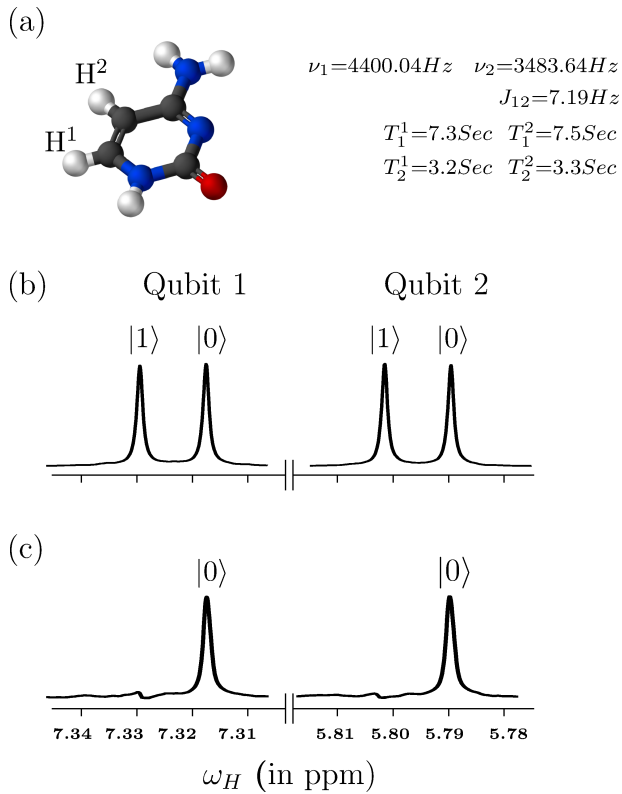}
\caption{(Color online) (a) Molecular
structure of cytosine with the two qubits labeled
as $H^{1}$ and $H^{2}$
and tabulated system parameters 
with chemical shifts $\nu_i$ and scalar coupling $J_{12}$ (in
Hz) and relaxation times $T_{1}$ and $T_{2}$ (in seconds)
(b) 
NMR spectrum obtained after a $\pi/2$
readout pulse on the thermal equilibrium state.
The resonance lines of each qubit are labeled by
the corresponding logical states of the other
qubit and
(c) NMR spectrum of the pseudopure $\vert 0 0 \rangle$
state.
}
\label{molecule}
\end{figure}
\subsection{NMR details}
The two protons of the molecule cytosine 
encode the two qubits.  The two-qubit molecular
structure, system parameters and pseudopure and thermal
initial states are shown in Figs.~\ref{molecule}(a)-(c).  
The Hamiltonian of a two-qubit system in the
rotating frame is given by
\begin{equation}
H = \sum_{i=1}^{2} \nu_i I_{iz} +  \sum_{i <
j, i=1}^{2} J_{ij} I_{iz}
I_{jz}
\end{equation}
where $\nu_i$ are the Larmor frequencies of the
spins and $J_{ij}$ is the spin-spin coupling
constant. An average longitudinal T$_1$ relaxation time
of 7.4 $s$ and an average transverse T$_2$ relaxation time of
3.25 $s$ was experimentally measured for both the
qubits.
All experiments were performed at an ambient temperature of
298 K on a Bruker Avance III 600 MHz NMR spectrometer equipped with a QXI
probe.  
The two-qubit system
was initialized into the pseudopure state $\vert 00 \rangle$
using the spatial averaging technique~\cite{cory-physicad},
with the
density operator given by
\begin{equation}
\rho_{00} = \frac{1-\epsilon}{4} I
+ \epsilon \vert 00 \rangle \langle 00 \vert
\label{ppure}
\end{equation}
with a thermal polarization $\epsilon \approx
10^{-5}$ and $I$ being a $4 \times 4$
identity operator.  The experimentally created
pseudopure state $\vert 00 \rangle$ was
tomographed with a fidelity of $0.99$.
The pulse propagators for selective excitation were
constructed using the GRAPE algorithm~\cite{tosner-jmr-09} to design
the amplitude and phase modulated RF profiles.  
Selective excitation was typically achieved with pulses of
duration 1 ms.
Numerically
generated GRAPE pulse profiles were optimized to be robust
against RF inhomogeneity and had an average fidelity of 
$ \ge 0.99$.
All experimental density matrices were reconstructed using a
reduced tomographic protocol~\cite{leskowitz-pra-04}, with
the set of operations given by $\{II, IX, IY, XX\}$ being
sufficient to determine all 15 variables for the two-qubit
system.  Here $I$ is the identity (do-nothing operation) and
$X (Y)$ denotes a single spin operator that can be
implemented by applying a corresponding spin selective
$\pi/2$ pulse. The fidelity of an experimental
density matrix was computed by 
measuring the
projection between the
theoretically expected and experimentally
measured states using the Uhlmann-Jozsa
fidelity measure~\cite{uhlmann-fidelity,jozsa-fidelity}:
\begin{equation}
F =
\left(Tr \left( \sqrt{
\sqrt{\rho_{\rm theory}}
\rho_{\rm expt} \sqrt{\rho_{\rm theory}}
}
\right)\right)^2
\label{fidelity}
\end{equation}

where $\rho_{\rm theory}$ and $\rho_{\rm expt}$ denote the
theoretical and experimental density matrices
respectively. 
\subsection{Super-Zeno for state preservation}
\label{statesection}
When the subspace $\cal P$ is a one-dimensional subspace,
and hence consists of a single state, the super-Zeno
scheme becomes a state preservation scheme.

\begin{figure}[hbtp] 
\centering
\includegraphics[angle=0,scale=1]{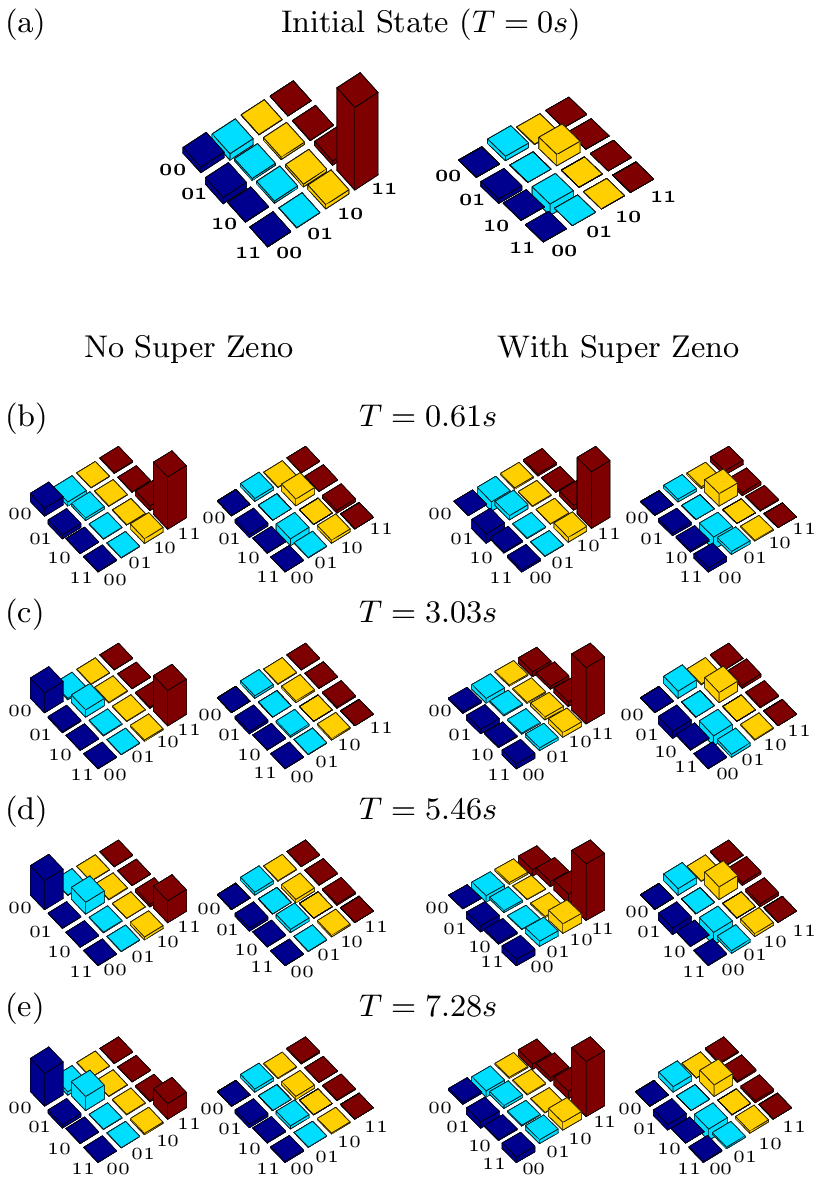} 
\caption{ (Color online)
Real (left) and imaginary (right) parts of the experimental
tomographs of the (a) $ \vert 11\rangle$ state, with a computed
fidelity of 0.99.  (b)-(e) depict the state at $T = 0.61,
3.03, 5.46, 7.28 s$, with the tomographs on the left and the
right representing the state without and after applying the
super-Zeno preserving scheme, respectively.  The rows and
columns are labeled in the computational basis ordered from
$\vert 00 \rangle$ to $\vert 11 \rangle$.
}
\label{11tomo}
\end{figure}

\noindent{\bf Product states:}
We begin by 
implementing the super-Zeno scheme on the product state $\vert
11 \rangle$ of two qubits, where the Hilbert space can be
decomposed as a direct sum of the
subspaces ${\cal P}=\{\vert 11 \rangle\}$
and ${\cal Q}=\{ \vert 00 \rangle, \vert 01 \rangle, \vert
10 \rangle)\}$. The super-Zeno pulse $\bf J$ to
protect the state $\vert 11 \rangle \in {\cal P}$  
is given by Eqn.~(\ref{Jeqn}):
\begin{equation}
{\bf J}= I - 2 \vert 11 \rangle \langle 11 \vert
\end{equation}
with the corresponding matrix form 
\begin{equation}
{\bf J}= \left(\begin{array}{cccc}
1&0&0&0 \\
0&1&0&0 \\
0&0&1&0 \\
0&0&0&-1
\end{array}
\right)
\end{equation}
\begin{figure}[hbtp]
\centering
\includegraphics[angle=0,scale=1.0]{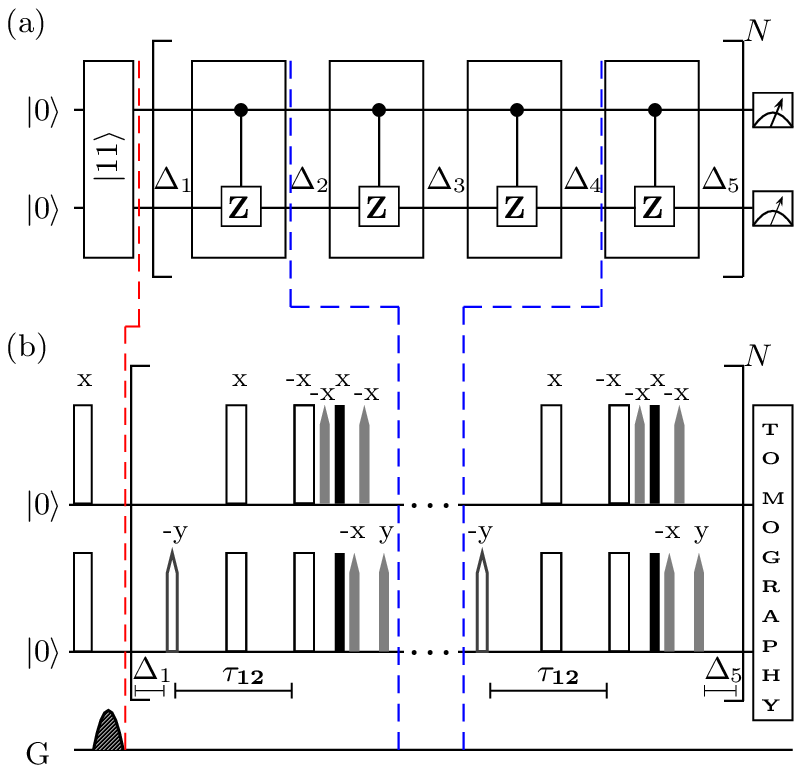}
\caption{ (Color online)
(a) Quantum circuit for preservation of the state $\vert 1 1
\rangle$ using the super-Zeno scheme.  
$\Delta_i = x_i t, (i=1...5)$
denote time intervals punctuating the unitary
operation blocks.  Each unitary operation block
contains a controlled-phase gate ($Z$), with the first (top) qubit 
as the control and the second (bottom) qubit as the
target. 
The entire scheme is repeated $N$ times
before measurement (for our experiments
$N=30$).  (b) Block-wise depiction of the
corresponding NMR pulse sequence. A $z$-gradient is applied
just before the super-Zeno pulses, to clean up 
undesired residual magnetization.  The unfilled 
and black rectangles represent
hard $180^{0}$ and $90^{0}$
pulses respectively, while
the unfilled  
and gray-shaded conical shapes represent
$180^{0}$ and $90^{0}$ pulses
(numerically optimized using GRAPE) respectively;
$\tau_{12}$ is
the evolution period under the $J_{12}$ coupling. Pulses are
labeled with their respective phases and unless explicitly
labeled, the phase of the pulses on the second (bottom)
qubit are the same as those on the first (top) qubit.
}
\label{11ckt}
\end{figure}
The super-Zeno circuit to preserve the $\vert 11 \rangle$
state, and the corresponding NMR pulse sequence is
given in Fig.~\ref{11ckt}. The controlled-phase gate
($Z$) in Fig.~\ref{11ckt}(a) which replicates the unitary kick ${\bf J}$ for
preservation of the $\vert 11 \rangle$ state is implemented
using a set of three sequential gates: two Hadamard gates
on the second qubit sandwiching a controlled-NOT gate
({\rm CNOT}$_{12}$), with the first qubit as the control
and the second qubit as the target. 
The $\Delta_i$ time interval in Fig.~\ref{11ckt}(a)
is given by $\Delta_i = x_i t$, with $x_i$ as
defined in Eqn.~(\ref{superzenoeqn}).
The five $\Delta_i$ time intervals were worked
to be $0.095 m$s, $0.25 m$s, $0.3 m$s, $0.25 m$s, 
and $0.095 m$s respectively, for $t=1 m$s.
One run of the super-Zeno circuit (with four inverting ${\bf J}$s
and five $\Delta_i$ time evolution periods) takes
approximately $300 m$s and 
the entire preserving sequence $W_N(t)$ 
in Eqn.~(\ref{superzenoeqn}) was applied
$30$ times.
The final state of the system was reconstructed using
state tomography and the real and imaginary
parts of the tomographed experimental density matrices without
any preservation and after applying the
super-Zeno scheme,
are shown in
Fig.~\ref{11tomo}. 
The initial $\vert 11 \rangle$ state (at time $T=0 s$)
was created (using the spatial 
averaging scheme) with a fidelity of 0.99.
The tomographs (on the right in
Fig.~\ref{11tomo}) clearly show that state evolution 
has been frozen with the super-Zeno scheme.

\noindent{\bf Entangled state preservation:}
We next apply the super-Zeno scheme to
preserve an entangled state in our system of two qubits.
We choose the singlet state
$\frac{1}{\sqrt{2}}(\vert 0 1 \rangle - \vert 1 0
\rangle)$ as the entangled state to be
preserved.  It is well known that
entanglement is an important but fragile computational
resource and constructing schemes to protect entangled
states from evolving into other states, is of considerable
interest in quantum information processing
~\cite{nielsen-book-02}. 

We again write the Hilbert space as a 
direct sum of two subspaces: the subspace being
protected and the subspace orthogonal to it.
In this case, the one-dimensional subspace 
${\cal P}$ being protected is  
\begin{equation}
{\cal P} = \left \{
\frac{1}{\sqrt{2}}\left(\vert 01 \rangle- \vert 10\rangle
\right) \right\}
\end{equation}
and the orthogonal subspace ${\cal Q}$ into which
one would like to prevent leakage is 
\begin{equation}
{\cal Q} = \left \{
\frac{1}{\sqrt{2}}(\vert 01 \rangle +\vert 10 \rangle),
\vert 00 \rangle, \vert 11 \rangle
\right \}
\end{equation}
\begin{figure}[hbtp]
\centering
\includegraphics[angle=0,scale=1.0]{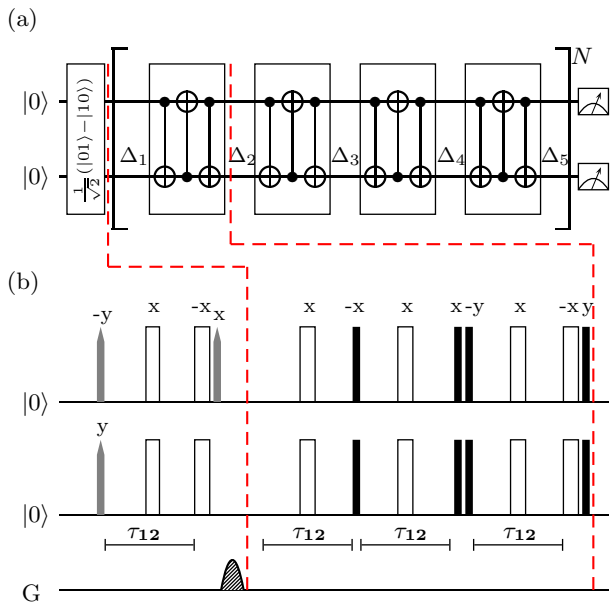}
\caption{ (Color online)
(a) Quantum circuit for preservation of the
singlet state using the super-Zeno scheme.  $\Delta_i,
(i=1...5)$ denote time intervals punctuating
the unitary operation blocks.  
The entire scheme is repeated $N$ times before
measurement (for our
experiments $N=10$).  (b) NMR pulse sequence corresponding to one
unitary block of the circuit in (a). 
A $z$-gradient is applied just
before the super-Zeno pulses, to clean up undesired
residual magnetization.  The unfilled rectangles represent hard
$180^{0}$ pulses, the black filled rectangles representing hard $90^{0}$ pulses,
while the shaded shapes represent numerically optimized (using GRAPE) pulses
and the gray-shaded shapes representing $90^{0}$ pulses respectively; 
$\tau_{12}$ is
the evolution period under the $J_{12}$ coupling. Pulses are
labeled with their respective phases and unless explicitly
labeled, the phase of the pulses on the second (bottom)
qubit are the same as those on the first (top) qubit.
}
\label{singletckt}
\end{figure}
The super-Zeno pulse to protect the singlet state 
as constructed using Eqn.~(\ref{Jeqn}) is:
\begin{equation}
{\bf J}= I -  \left(
\vert 01\rangle \langle01 \vert+ \vert 10
\rangle\langle10 \vert- \vert 01\rangle\langle10
\vert- \vert 10\rangle\langle 01\vert \right)
\end{equation}
with the corresponding matrix form:
\begin{equation}
{\bf J}= \left(\begin{array}{cccc}
1&0&0&0 \\
0&0&1&0 \\
0&1&0&0 \\
0&0&0&1
\end{array}
\right)
\end{equation}

\begin{figure}[hbtp]
\centering
\includegraphics[angle=0,scale=1.0]{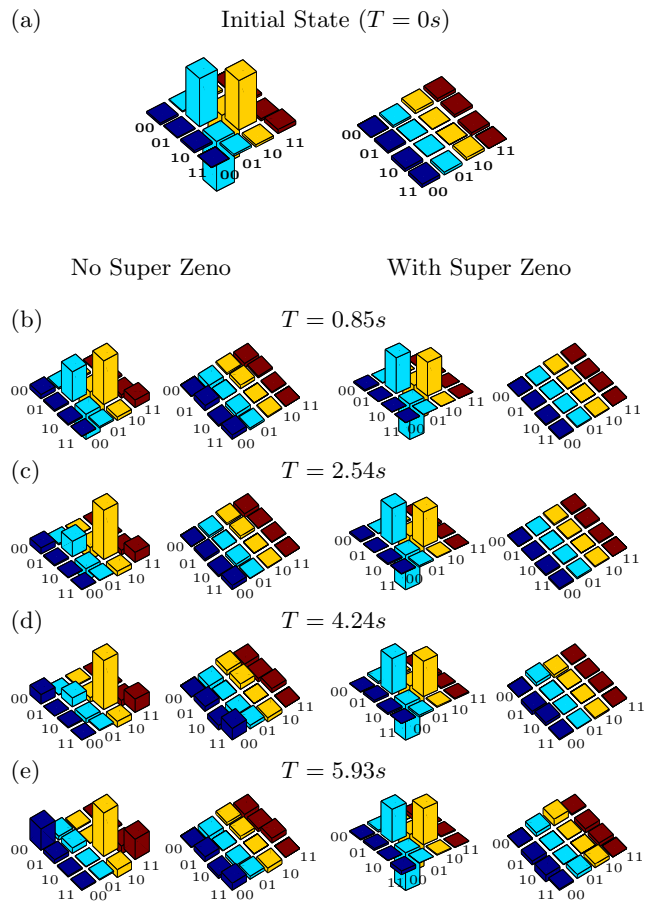}
\caption{ (Color online)
Real (left) and imaginary (right) parts of the experimental
tomographs of the (a) $\frac{1}{\sqrt{2}}(\vert 01\rangle
-\vert 1 0 \rangle)$ (singlet) state, with a computed
fidelity of 0.99.  (b)-(e) depict the state at $T = 0.85,
2.54, 4.24, 5.93 s$, with the tomographs on the left and the
right representing the state without and after applying the
super-Zeno preserving scheme, respectively.  The rows and
columns are labeled in the computational basis ordered from
$\vert 00 \rangle$ to $\vert 11 \rangle$.
}
\label{singlettomo}
\end{figure}

The quantum circuit and the NMR
pulse sequence for preservation of the 
singlet state using the super-Zeno scheme are given in
Fig.~\ref{singletckt}. Each {\bf J} inverting pulse
in the unitary block in the circuit is decomposed
as a sequential operation of three non-commuting controlled-NOT
gates:~{\rm CNOT$_{12}$-CNOT$_{21}$-CNOT$_{12}$},
where {\rm CNOT$_{ij}$} denotes a controlled-NOT with $i$
as the control and $j$ as the target qubit. 
The five $\Delta_i$ time intervals were worked
to be $0.95 m$s, $2.5 m$s, $3 m$s, $2.5 m$s, 
and $0.95 m$s respectively, for $t = 10m$s.
One run of the super-Zeno circuit (with four inverting ${\bf J}$s
and five $\Delta_i$ time evolution periods) takes
approximately $847 m$s
and the entire super-Zeno preserving sequence 
$W_N(t)$ 
in Eqn.~(\ref{superzenoeqn}), is applied
$10$ times.
The singlet state was prepared from an initial
pseudopure state $\vert 00 \rangle$ by a sequence
of three gates: a 
non-selective NOT gate (hard $\pi_x$ pulse) on both qubits, 
a Hadamard gate and a {\rm CNOT}$_{12}$ gate. The 
singlet state thus prepared
was computed to have a fidelity of 0.99.
The effect of chemical shift evolution during the delays was
compensated for with refocusing pulses.
The final singlet state has been reconstructed using
state tomography, and the real and imaginary
parts of the tomographed experimental density matrices without
any preservation and after applying the
super-Zeno scheme,
are shown in
Fig.~\ref{singlettomo}. As can be seen from the
experimental tomographs in Fig.~\ref{singlettomo}, the evolution of the
singlet state is almost completely frozen by the
super-Zeno sequence upto nearly 6 $s$, while without
any preservation the state has leaked into the
orthogonal subspace within 2 $s$. 

\begin{figure}[hbtp] 
\includegraphics[angle=0,scale=1.0]{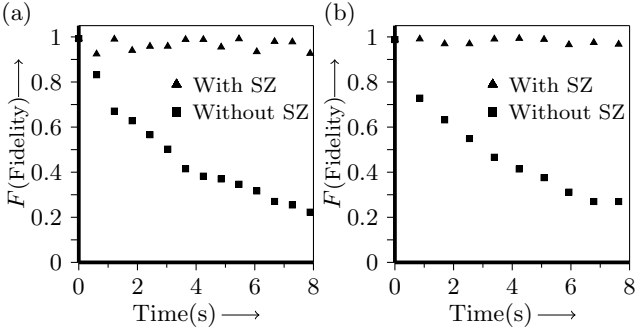} 
\caption{
Plot of fidelity versus time of (a) the $\vert 11\rangle$
state and (b) the $\frac{1}{\sqrt{2}}(\vert 0 1 \rangle -
\vert 1 0 \rangle)$ (singlet) state, without any preserving
scheme and after the super-Zeno preserving sequence. The
fidelity of the state with the super-Zeno preservation
remains close to 1.
}
\label{fidelityfig}
\end{figure}

\noindent{\bf Estimation of state fidelity:}
The plots of state fidelity 
versus time are shown in
Fig.~\ref{fidelityfig} for the state $\vert 11 \rangle$ and
the singlet state, with and without the super-Zeno
preserving sequence.  
The deviation density matrix is renormalized
at every point 
and the state
fidelity 
is estimated 
using the definition in Eqn.~(\ref{fidelity}).
Renormalization is performed since our
focus 
here is on the quantum state
of the spins contributing to the signal 
and not in the number per se of participating
spins~\cite{brazil-norm}.
The plots in Fig.~\ref{fidelityfig} and the tomographs
in Fig.~\ref{11tomo} and Fig.~\ref{singlettomo} show that
with super-Zeno protection, the state remains confined
to the $\vert 11 \rangle$ (singlet) part of the density
matrix, while without the protection scheme, the state
leaks into the orthogonal subspace.
As seen from both plots in Fig.~\ref{fidelityfig}, the
state evolution of specific states can be arrested
for quite a long time using the super-Zeno preservation
scheme, while leakage probability of the state to other
states in the orthogonal subspace spanned by ${\cal Q}$ is
minimized. A similar renormalization procedure is adopted in the 
subsequent sections where we plot the leak fraction and
entanglement parameters(Fig.~\ref{leakagefig}
and Fig.~\ref{entangfig}).
\subsection{Super-Zeno for subspace preservation}
\label{subspacesection}
\begin{figure}
\centering
\includegraphics[angle=0,scale=1.0]{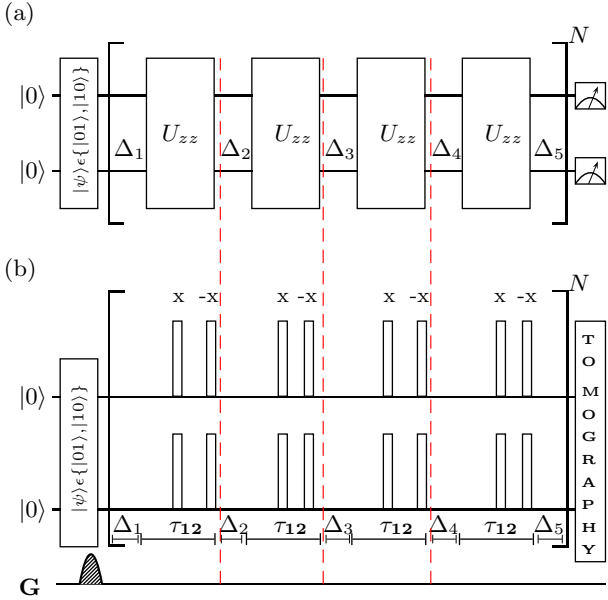}
\caption{ (Color online)
(a) Quantum circuit for preservation of the $\{01, 10\}$
subspace using the super-Zeno scheme.  $\Delta_i, (i=1...5)$
denote time intervals punctuating the unitary
operation blocks.  
The entire scheme is repeated $N$ times
before measurement (for our experiments
$N=30$).  (b) NMR pulse sequence corresponding to
the circuit in (a).  A $z$-gradient is applied just before
the super-Zeno pulses, to clean up undesired
residual magnetization.  The unfilled rectangles represent
hard $180^{0}$ pulses;
$\tau_{12}$ is the evolution period under the $J_{12}$
coupling. Pulses are labeled with their respective phases.
}
\label{subspaceckt}
\end{figure}
While in the previous subsection, the super-Zeno scheme was shown to be
effective in arresting the evolution of 
a one-dimensional subspace (as applied to
the cases of a product and an entangled state),
the scheme is in fact more general.
For example, if we choose a two-dimensional subspace in the
state space of two qubits and protect it by the
super-Zeno scheme, then any state in this subspace
is expected to remain within this subspace and not
leak into the orthogonal subspace. While the state
can meander within this subspace, its evolution out of the
subspace is frozen.

We now turn to implementing 
the super-Zeno scheme for subspace preservation, by
constructing the {\bf J} operator to preserve a
general state embedded in a two-dimensional subspace.
We choose the subspace  
spanned by ${\cal P}=\{\vert 01\rangle, \vert 10\rangle\}$
as the subspace to be preserved, with its orthogonal
subspace now being 
${\cal Q}=\{\vert 00 \rangle, \vert 11 \rangle\}$.
It is worth noting that within the subspace being
protected, we have product as well as entangled states.

The super-Zeno pulse {\bf J} to protect a general state 
$\vert \psi\rangle \in P$ can be constructed as:
\begin{equation}
{\bf J}= I-2 \left( 
\vert01\rangle\langle01\vert-\vert10\rangle\langle10\vert 
\right)
\end{equation}

with the corresponding matrix form 
\begin{equation}
{\bf J}= \left( \begin{array}{cccc}
1&0&0&0 \\
0&-1&0&0 \\
0&0&-1&0 \\
0&0&0&1
\end{array}
\right)
\end{equation}

\begin{figure}[t]
\centering
\includegraphics[angle=0,scale=1.0]{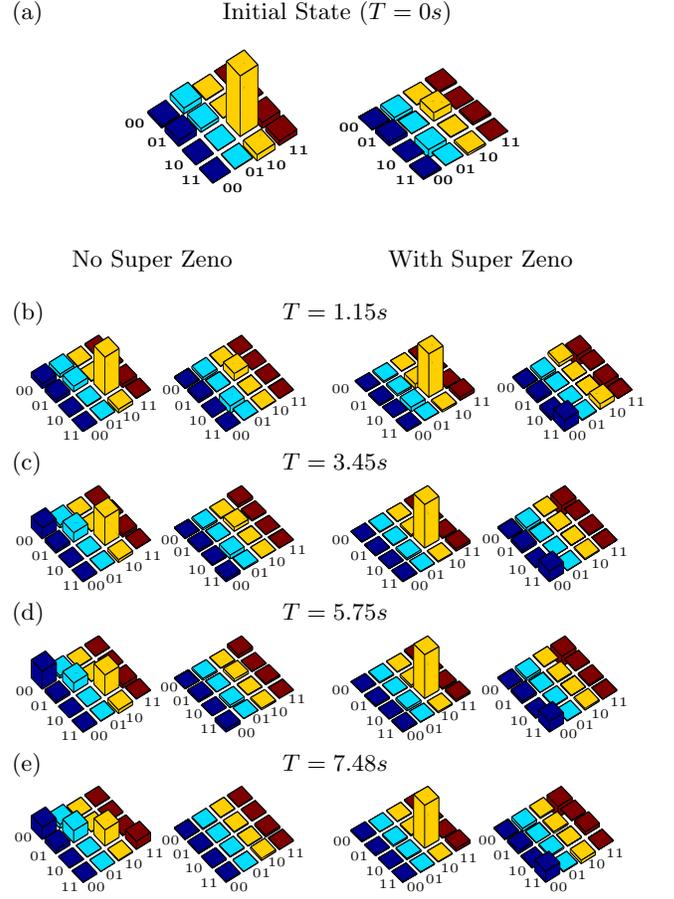}
\caption{ (Color online)
Real (left) and imaginary (right) parts of
the experimental tomographs of the (a) $\vert 1 0 \rangle$ 
state in the two-dimensional subspace $\{ 01, 10 \}$,
with a computed fidelity of 0.98.
(b)-(e) depict the state at $T = 1.15,
3.45, 5.75, 7.48 s$, with the tomographs on the left and the
right representing the state without and after applying the
super-Zeno preserving scheme, respectively.  The rows and
columns are labeled in the computational basis ordered from
$\vert 00 \rangle$ to $\vert 11 \rangle$.
}
\label{10tomo}
\end{figure}
The quantum circuit and corresponding NMR pulse sequence to
preserve a general state in the $\{\vert 01 
\rangle, \vert 10 \rangle \}$ subspace is
given in Fig.~\ref{subspaceckt}. The unitary kick (denoted
as $U_{zz}$ in the unitary operation block in
Fig.~\ref{subspaceckt}(a)) is  
implemented by  tailoring the gate time to the $J$-coupling
evolution interval of the system Hamiltonian, sandwiched by
non-selective $\pi$ pulses (NOT gates), to refocus undesired
chemical shift evolution during the action of the
gate. 
The five $\Delta_i$ intervals were worked
to be $0.95 m$s, $2.5 m$s, $3 m$s, $2.5 m$s 
and $0.95 m$s respectively, for $t = 10 m$s.
One run of the super-Zeno circuit (with four inverting ${\bf J}$s
and five $\Delta_i$ time evolution periods) takes
approximately $288 m$s
and the entire super-Zeno preserving sequence 
$W_N(t)$ in
Eqn.~(\ref{superzenoeqn}), is applied
$30$ times.
\begin{figure}
\centering
\includegraphics[angle=0,scale=1.0]{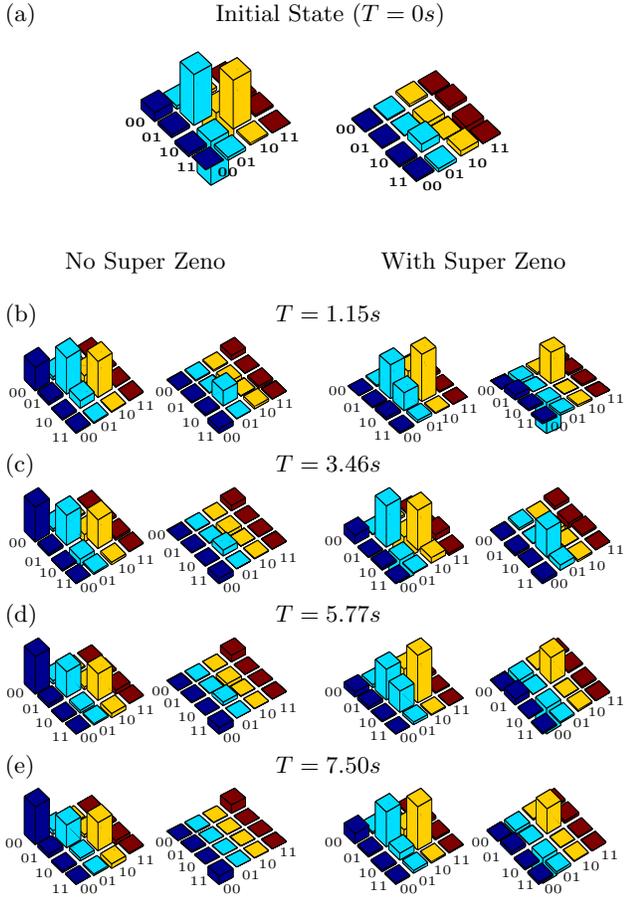}
\caption{ (Color online) Real (left) and imaginary (right) parts of
the experimental tomographs of the (a) 
$\frac{1}{\sqrt{2}}(\vert 0 1 \rangle - \vert 1 0 \rangle)$
(singlet) 
state in the two-dimensional subspace $\{ 01, 10 \}$,
with a computed fidelity of 0.98.
(b)-(e) depict the state at $T = 1.15,
3.46, 5.77, 7.50 s$, with the tomographs on the left and the
right representing the state without and after applying the
super-Zeno preserving scheme, respectively.  The rows and
columns are labeled in the computational basis ordered from
$\vert 00 \rangle$ to $\vert 11 \rangle$.
}
\label{singletsubspacetomo}
\end{figure}


\noindent{\bf Preservation of  product states in
the subspace:}
We implemented the subspace-preserving scheme on
two different (separable) states $\vert 0 1 \rangle$ and
$\vert 1 0 \rangle$ in the subspace ${\cal P}$.
The efficacy of the preserving unitary is
verified by tomographing the experimental density
matrices at different time points and computing the
state fidelity. Both the $\vert 0 1 \rangle$ and
$\vert 1 0 \rangle$ states remain within the
subspace ${\cal P}$ and
do not leak out to the orthogonal subspace ${\cal Q} =
\{ \vert 00 \rangle , \vert 11 \rangle \}$. 

The final $\vert 1 0 \rangle$ state has been reconstructed using
state tomography, and the real and imaginary
parts of the experimental density matrices without
any preservation and after applying the
super-Zeno scheme,
tomographed at different time points, are shown in
Fig.~\ref{10tomo}. As can be seen from the
experimental tomographs, the evolution of the
$\vert 1 0 \rangle$ state out of the
subspace is almost completely frozen by the
super-Zeno sequence upto nearly 7.5 $s$, while without
any preservation the state has leaked into the
orthogonal subspace within 3.5 $s$. 
The tomographs for the $\vert 0 1 \rangle$ state show a
similar level of preservation (data not shown).


\noindent{\bf Preservation of an entangled state in 
the subspace:}
We now prepare an entangled state (the singlet state)
embedded in the two-dimensional ${\cal P}=\{\vert 01
\rangle, \vert 10 \rangle \}$
subspace, and used the subspace-preserving scheme described
in Fig.~\ref{subspaceckt} to protect ${\cal P}$.
The singlet state was reconstructed using
state tomography, and the real and imaginary
parts of the tomographed experimental density matrices without
any preservation and after applying the
super-Zeno scheme,
are shown in
Fig.~\ref{singletsubspacetomo}. 
As can be seen from the experimental
tomographs, the state evolution remains
within the ${\cal P}$ subspace but the state
itself does not remain maximally entangled. 

\begin{figure}[t]
\begin{center}
\includegraphics[angle=0,scale=1.0]{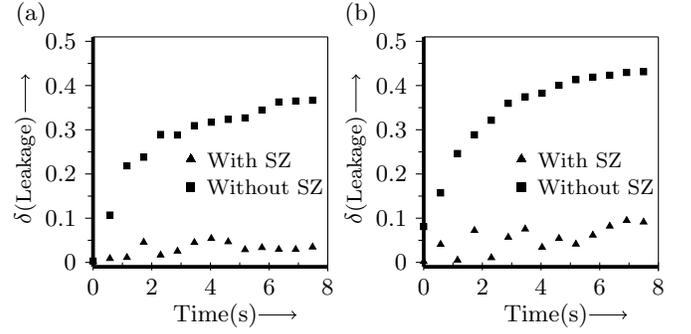}
\caption{Plot of leakage fraction 
from the $\{\vert 01\rangle, \vert 10\rangle \}$ subspace to its orthogonal
subspace $\{\vert 00\rangle, \vert 11\rangle \}$ of (a) the $\vert 1 0 \rangle$
state and (b) the
$\frac{1}{\sqrt{2}}(\vert 0 1 \rangle - \vert 1 0 \rangle)$
(singlet) state, without any preservation and after
applying the super-Zeno sequence. The leakage to the
orthogonal subspace is minimal (remains close to zero)
after applying the super-Zeno scheme.
\label{leakagefig}
}
\end{center}
\end{figure} 
\noindent{\bf Estimating leakage outside subspace:}
The subspace-preserving capability of the circuit
given in Fig.~\ref{subspaceckt} was quantified by
computing a leakage parameter that defines the amount of
leakage of the state to the orthogonal
${\cal Q}=\{\vert 00 \rangle, \vert 11 \rangle \}$ subspace. 
For a given density operator $\rho$ the ``leak
(fraction)'' $\delta$, into the subspace ${\cal Q}$ is defined 
as
\begin{equation}
\delta = 
\langle 00 \vert \rho \vert 00 \rangle + 
\langle 11 \vert \rho \vert 11 \rangle  
\end{equation}
The leak (fraction) $\delta$
versus time is plotted in Figs.~\ref{leakagefig}(a) and (b), for
the $\vert 1 0\rangle$ and the singlet state respectively, with
and without applying the super-Zeno subspace-preserving
sequence. The leakage parameter remains close to zero
for both kinds of states, proving the success 
and the generality of the
super-Zeno scheme.
\subsection{Preservation of entanglement}
The amount of entanglement that remains in the state
after a certain time is quantified by
an entanglement parameter denoted by $\eta$.
Since we are dealing with mixed bipartite
states of two qubits, all entangled states
will be negative under partial
transpose (NPT).  
For such NPT states, a reasonable measure of
entanglement is the minimum eigen value of 
the partially transposed density operator.
For a given experimentally tomographed 
density operator $\rho$,
we obtain
$\rho^{PT}$ by taking a partial transpose
with respect to one of the qubits.
The entanglement parameter $\eta$ 
for the state $\rho$ in terms of the
smallest eigen value $E^{\rho}_{\rm Min}$  
of $\rho^{PT}$ is defined as 
\begin{equation}
\eta = \left\{\begin{array}{ll}
-E^{\rho}_{\rm Min} &{\rm if~} E^{\rho}_{\rm Min} <0
\\
&\\
\phantom{-}0   &{\rm if~} E^{\rho}_{\rm Min} > 0
\end{array}\right.
\end{equation}
We will use this entanglement parameter $\eta$ to
quantify the amount of entanglement at different
times.

The maximally entangled singlet state was created
and its evolution studied in two different
scenarios. In the first scenario described
in Sec.~\ref{statesection}, the 
singlet state was protected against evolution by the
application of the super-Zeno scheme. In the second
scenario described
in Sec.~\ref{subspacesection}, a two-dimensional subspace containing
the singlet state was protected using the super-Zeno
scheme. For the former case, one expects that the state will
remain a singlet state, while in the latter case, it
can evolve within the protected two-dimensional subspace.
Since in the second case, the protected subspace contains
entangled as well as separable states, one does not expect
preservation of entanglement to the same extent as expected
in the first case, where the one-dimensional subspace
defined by the singlet state itself is protected.
The experimental tomographs at different times and
fidelity for the case of state protection and 
the leakage fraction for the case of subspace protection  
have been discussed in detail in the previous subsections.
\begin{figure}[h]
\centering
\includegraphics[angle=0,scale=1.0]{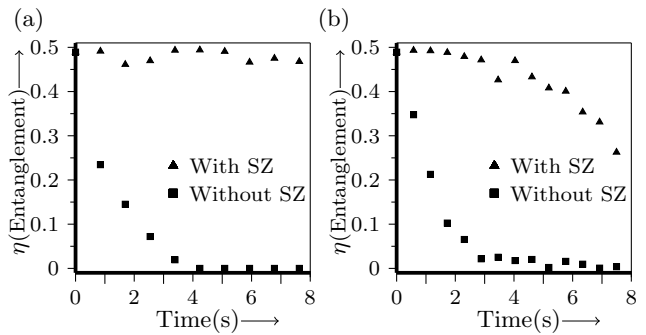}
\caption{Plot of entanglement parameter $\eta$ with time,
with and without applying the super-Zeno sequence,
computed for (a) the
$\frac{1}{\sqrt{2}}(\vert 0 1 \rangle - \vert 1 0 \rangle)$
(singlet) state, and (b) the same singlet state when
embedded in the subspace $\{\vert 01 \rangle, 
\vert 10 \rangle \}$ being
preserved.}
\label{entangfig}
\end{figure} 
Here we focus our attention on the entanglement present in
the state at different times.
The entanglement parameter $\eta$ for the
evolved singlet state 
is plotted as
a function of time and is
shown in Figs.~\ref{entangfig}(a) and (b),
after 
applying the state-preserving and the subspace-preserving
super-Zeno sequence respectively. 
In both cases,
the state becomes disentangled very quickly (after
approximately 2 $s$) if no super-Zeno preservation
is performed. After applying the state-preserving
super-Zeno sequence (Fig.~\ref{entangfig}(a)), the
amount of entanglement in the state remains 
close to maximum for a long
time (upto 8 $s$). After applying the
subspace-preserving super-Zeno sequence
(Fig.~\ref{entangfig}(b)), the state shows some
residual entanglement over long times but it is
clear that the state is no longer maximally
entangled. This implies that 
the subspace-preserving sequence does not
completely preserve the entanglement of the singlet state,
as expected.
However, while the singlet state becomes mixed over time,
its evolution
remains confined to states within the 
two-dimensional subspace (${\cal P} = \{\vert 
01 \rangle, \vert 10 \rangle \}$) 
being preserved as is shown in Fig.~\ref{leakagefig},
where we calculate the leak (fraction). 
\section{Concluding Remarks}
\label{concl}
In summary, we have experimentally demonstrated that the
super-Zeno scheme can efficiently preserve states in one-
and two-dimensional subspaces, by preventing leakage to
a subspace orthogonal to the subspace being preserved.  We have
implemented the super-Zeno sequence on product as well as on
entangled states, embedded in one- and two-dimensional
subspaces of a two-qubit NMR quantum information processor.

We emphasize here that the strength of the super-Zeno
protection scheme lies in its ability to preserve the state
such that while the number of spins in that particular
state reduces with time, the state remains the same.
Without the super-Zeno protection, the number of spins in the state
reduces with time and the state itself migrates towards
a thermal state, reducing the fidelity.
Our work adds to the arsenal of real-life attempts to
protect against evolution of states in quantum computers and
points the way to the possibility of developing hybrid
strategies (combining the super-Zeno scheme with other
schemes such as dynamical decoupling sequences) to tackle
preservation of fragile computational resources such as
entangled states.

\begin{acknowledgments}
All experiments were performed on a Bruker Avance-III 600
MHz FT-NMR spectrometer at the NMR Research Facility at
IISER Mohali.  HS is funded by a Government of India
CSIR-NET JRF fellowship.
\end{acknowledgments}
\end{document}